\documentclass[sigconf,screen]{acmart}

\usepackage{graphicx}
\usepackage{setspace}               
\usepackage{multicol}               
\usepackage{xcolor}
\usepackage{caption}
\usepackage{subcaption}
\usepackage{csquotes}

\usepackage[utf8]{inputenc}
\usepackage{url}
\usepackage{hyperref}
\usepackage{amsfonts}
\usepackage{float} 
\usepackage{listings}
\usepackage{color}
\usepackage{bibentry} 
\usepackage{enumitem}
\usepackage{multirow}
\usepackage{geometry}
\usepackage{dirtytalk}
\usepackage{tabularx}
\usepackage{balance}

\usepackage[english]{babel}
\usepackage{soul}

\definecolor{edit}{HTML}{000000}

\AtBeginDocument{%
  \providecommand\BibTeX{{%
    \normalfont B\kern-0.5em{\scshape i\kern-0.25em b}\kern-0.8em\TeX}}}

\copyrightyear{2025}
\acmYear{2025}
\setcopyright{cc}
\setcctype{by}
\acmConference[C\&C '25]{Creativity and Cognition}{June 23--25, 2025}{Virtual, United Kingdom}
\acmBooktitle{Creativity and Cognition (C\&C '25), June 23--25, 2025, Virtual, United Kingdom}\acmDOI{10.1145/3698061.3726924}
\acmISBN{979-8-4007-1289-0/2025/06}

\begin{document}

\title{Beyond Productivity: Rethinking the Impact of Creativity Support Tools}

\author{Samuel Rhys Cox}
\email{srcox@cs.aau.dk}
\orcid{0000-0002-4558-6610}
\affiliation{%
  \institution{Aalborg University}
  \city{Aalborg}
  \country{Denmark}
}

\author{Helena Bøjer Djernæs}
\email{hbd@cs.aau.dk}
\orcid{0009-0003-9457-811X}
\affiliation{%
  \institution{Aalborg University}
  \city{Aalborg}
  \country{Denmark}
}

\author{Niels van Berkel}
\email{nielsvanberkel@cs.aau.dk}
\orcid{0000-0001-5106-7692}
\affiliation{%
  \institution{Aalborg University}
  \city{Aalborg}
  \country{Denmark}
}

\begin{abstract}

Creativity Support Tools (CSTs) are widely used across diverse creative domains, with generative AI recently increasing the abilities of CSTs. To better understand how the success of CSTs is determined in the literature, we conducted a review of outcome measures used in CST evaluations. Drawing from (\textit{n=173}) CST evaluations in the ACM Digital Library, we identified the metrics commonly employed to assess user interactions with CSTs. Our findings reveal prevailing trends in current evaluation practices, while exposing underexplored measures that could broaden the scope of future research. Based on these results, we argue for a more holistic approach to evaluating CSTs, encouraging the HCI community to consider not only user experience and the quality of the generated output, but also user-centric aspects such as self-reflection and well-being as critical dimensions of assessment. 
We also highlight a need for validated measures specifically suited to the evaluation of generative AI in CSTs.
\end{abstract}



\begin{teaserfigure}
 \centering
 \includegraphics[width=1\textwidth]{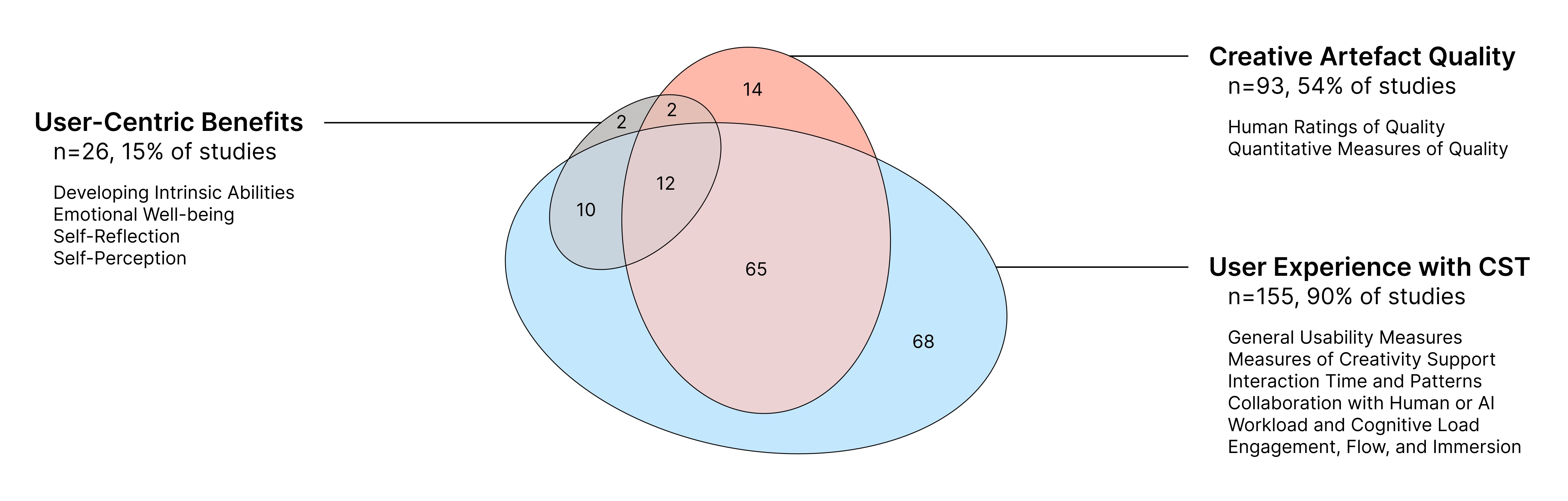}
 \caption{\textcolor{edit}{Measures used in (\textit{n=173}) empirical studies of CSTs from a survey of 10 years of ACM DL literature. Measures of \textit{User Experience with CSTs} were most prevalent (90\%), followed by measures of \textit{Creative Artefact Quality} (54\%), and measures of \textit{User-Centric Benefits} were least prevalent (15\%).}}
 \Description{Venn diagram illustrating the overlap between three categories of outcome measures in CST studies: User Experience, Creative Artefact Quality, and User-Centric Benefits. Most studies fall within User Experience alone or overlap with Creative Artefact Quality. Only a small number include User-Centric Benefits, with very few addressing all three. Labels note example measures used in each category.}
 \label{fig:teaser}
\end{teaserfigure}


\maketitle

\section{Introduction}

Creativity support tools (CSTs) have been widely adopted to scaffold, guide, and enhance the creation of creative artefacts. 
CSTs have been applied across diverse domains, including creative writing and storytelling~\cite{zarei2020investigating,suh2022privacytoon}, idea exploration~\cite{kim2022mixplorer,marinussen2019being}, and product design~\cite{kato2017f3,zhang2024exploring}, to name just a few examples.
However, due to the mixed nature of CSTs, where their use is both artistic~\cite{wolf2019paint} and focused on generation of content~\cite{girotto2019crowdmuse}, the evaluation of such tools can be broad and varied in aim.
The rapid advancement of generative AI has further complicated this landscape, prompting reflection on the evolving role of the human user in CST interactions~\cite{biermann2022tool}. 
As these technologies continue to grow in capability and accessibility, a pressing question emerges: should there be a philosophical shift in how the outcomes of CSTs are conceptualised and measured?


Motivated by this, we analyse 10 years of literature from the ACM Digital Library, surveying outcomes reported in empirical user studies of CSTs.
\textcolor{edit}{From a qualitative analysis of 173 papers, we} separate these outcomes into themes of: 
(1) measures of \textit{user experience} while using CSTs, (2) measures of \textit{creative artefact quality}, and (3) \textit{user-centric} measures of benefits to the \textit{individual themselves} beyond the generation of the creative artefact (such as learning or well-being).
As a result, we find that the majority of HCI CST research has focused on measures of user experience and creative artefact quality, 
with a paucity of studies measuring user-centric benefits.

From this, we generate a call to researchers and practitioners to focus on outcome measures that help the \textit{user themselves} through the creative process, as an area of focus in need of added attention. 
Additionally, we highlight need for validated user experience measures to account for growing affordances and limitations of using generative AI within CSTs.
While we acknowledge the importance of improving the quality of creative output, in this work we wish to draw attention to outcomes that improve the quality of life of those using CSTs.




\section{Related Work}

\textcolor{edit}{
Creativity is commonly defined as the generation of artefacts that are both novel (i.e., perceived as unique or original) and useful (i.e., perceived as valuable or feasible) within their given context~\cite{plucker2004isn}.
Creativity involves cognitive processes such as divergent thinking (the generation of multiple varied ideas), and convergent thinking (refining and selecting the most appropriate idea).
}

Building on this understanding of creativity, researchers in HCI have long explored how computational systems can support and enhance creative processes.
Creativity support tools (CST) have a long history in HCI literature \cite{frich2018twenty,frich2019mapping,remy2020evaluating}, and have gained renewed prominence with the increasing accessibility of AI technologies~\cite{pang2025understanding}.
Within this study, we define CSTs as systems or strategies that can be used to enhance and assist creative tasks, such as exploration, ideation, or creation. 
These include digital tools such as Fujinami et al.’s CST for painting with physical objects~\cite{fujinami2018painting}, Zhang et al.’s argumentative writing support tool~\cite{zhang2023visar}, and Ford et al.’s music composition tool for professional composers~\cite{ford2024reflection}.
CSTs also encompass non-digital support tools, as explored by Lundquist et al. \cite{lundqvist2018physical} in their study of how physical tools (e.g., pen and paper), compare to digital and hybrid ideation setups, from which they argue for a combination of both.

Providing relevant creativity support is highly dependent on the intended end-user. Many studies focus on how to support adults during creative processes, such as writing~\cite{schmitt2021characterchat,sultanum2021leveraging,gonccalves2017understanding}, drawing~\cite{hoffmann2024thermalpen,lawton2023drawing,pyakurel2023cogcues}) or ideation~\cite{park2023exergy,rezwana2022understanding,jeon2021fashionq}, and have done so both targetting beginners seeking to get started in a new creative field, as well as practitioners looking to enhance their capability and productivity. 
For instance, Hu et al.~\cite{hu2021exploring} investigated creativity support through a collaborative environment where an adult user and a robot took turns creating art using tangram pieces. The study measured dimensions of creativity such as flexibility, originality, and elaboration to assess the extent to which the system supported creative output. After the task, participants completed a questionnaire and took part in interviews to provide further insights into perceived helpfulness and usability. 
In contrast, studies focusing on supporting children’s creative processes often emphasise learning-oriented contexts~\cite{williams2024doodlebot,hagen2023evaluating}. 
For example, Yan et al.~\cite{yan2023nacanva}, developed a digital collage tool (for collecting and inspecting new aspects of nature) that encourages children to explore, learn and create with nature.

\textcolor{edit}{
Beyond the learning outcomes of CST studies discussed above, engaging in creative activities has been shown to offer significant benefits for the user themselves.
Fancourt and Finn~\cite{fancourt2020evidence} offer an extensive scoping review that was published by the World Health Organization.
Based on a coverage of over 3000 studies they found:
``\textit{a major role for the arts in the prevention of ill health, promotion of health, and management and treatment of illness across the lifespan}''~\cite{fancourt2020evidence}, with the review discussing benefits from creative activities such as creative writing, dance, and painting.
More specifically, creative activities have been found to benefit learning~\cite{gajda2017creativity,KAPOOR2025101751}, emotional well-being~\cite{Conner04032018,Kaimal04052017}, mental health~\cite{fancourt2020evidence}, mindfulness~\cite{Curry01012005}, self-efficacy~\cite{Kaimal04052017,chandler1999creative}, and self-esteem~\cite{chandler1999creative}.
For example, brief (45 minute) free art-making facilitated by an art therapist improved both people's emotions and self-efficacy~\cite{Kaimal04052017}.
A creative writing program was found to enhance both the self-efficacy and self-esteem of adolescents~\cite{chandler1999creative}.
Finally, the type of creative activity has also been found to affect the extent of positive benefits, with Curry and Kasser~\cite{Curry01012005} finding that structured colouring activities (i.e., colouring in a mandala) led to greater reductions in anxiety compared to free colouring.
}

\textcolor{edit}{
Drawing this together, this review explores the outcome measures used in the evaluation of CSTs in prior work.
Importantly, we differentiate our work from that of Remy et al.~\cite{remy2020evaluating}, who surveyed the methodological approaches (e.g., surveys, interviews, and observations) used in CST evaluations up to 2019.
In contrast, our review focuses on the explicit outcome measures employed, examining the specific scales, metrics, and their prevalence across the literature.
}

\section{Methodology}
\textcolor{edit}{
Our approach was influenced by prior literature reviews within HCI~\cite{rapp2021human,zheng2022ux}. Specifically, we followed the four stages of: (1) Define: define the exclusion and inclusion criteria; (2) Search: develop search query and search relevant source; (3) Selection: filter resulting papers based on our selection and exclusion criteria; and (4) Analysis: perform qualitative analysis (open coding) on the final 173 total included papers
}

\subsection{Definition}
In this section, we define the inclusion and exclusion criteria applied during the selection process, to filter out relevant research papers from the initial 1241 included papers.

\subsubsection{Inclusion criteria}
This literature review includes studies that empirically evaluate user interactions with CSTs.

Beyond this, we include studies that integrate creative interaction and evaluate creativity-related outcomes, even when creativity is not the \textit{primary} focus of the tool. For example, `\textit{SelVReflect}'~\cite{wagener2023selvreflect} employs a guided 3D drawing process in virtual reality to support creative self-expression as part of a reflective practice. 

Additionally, to allow for a more broad surveying of creativity literature \textcolor{edit}{(and thereby a more holistic view of measures used)}, we included studies that use non-digital creativity tools, or that use forms of feedback to facilitate human-human creativity tasks. For example, Ma et al. developed an interface that visualises the gaze of another user during a drawing task on a shared canvas \cite{ma2025eyedraw}.

As the incorporation of AI into CSTs has increased, there have been competing definitions surrounding CSTs. While some prior work differentiates ``\textit{human-AI co-creativity}'' from CSTs (due to AI being framed as a partner to creativity rather than a tool~\cite{rezwana2022understanding}), we follow the definition of prior work that considered human-AI co-creativity as a form of CST~\cite{ghajargar2022redhead,biermann2022tool}, and the finding that CSTs could form part of a support network to creative output~\cite{chung2022artist}.
That is to say, co-creative AI can be seen to facilitate ideation and provides iterative feedback aligning with CST principles.


\subsubsection{Exclusion criteria}
Studies were excluded from the literature review, based on whether and how they evaluated a CST. 

We excluded studies of tools that automatically generated content that was not subsequently edited by the user. 
However, we included studies if the CST automatically generated some forms of content that was then editable and allowed for user interactions (e.g., generation and editing of webpage layouts \cite{buschek2020paper2wire}).

Likewise, we excluded studies that developed taxonomies of user groups (i.e., categories for how users behave, rather than explorations of how people interact with a specific support tool). For example, Xu et al. generated four different patterns of behaviour used when people use search engines to help with creativity tasks~\cite{xu2024idea}, and Palani et al. generated roles that people follow when using generative AI~\cite{palani2024evolving}.

We also excluded studies where creativity was primarily influenced through cognitive prompts rather than interactive or iterative scaffolding. For example, Yen et al.~\cite{yen2017listen} examined how different reflection conditions affected iterative design revisions, but did not introduce or evaluate a creativity support tool (CST). While reflection can influence creativity, it does not provide structured, external scaffolding or interactive support, distinguishing it from CSTs that actively guide or modify creative outputs.

Studies that include surveys of how people interact with and perceive existing CSTs were also excluded (e.g., \cite{chang2023prompt}), as they do not evaluate a specific tool within the study context. This includes online surveys reporting general usage trends of tools without presenting tool-specific evaluations.

We excluded papers that assessed interventions influencing creativity rather than evaluating a creativity support tool (CST) designed to support creativity directly. For example, Elgarf et al.~\cite{elgarf2022and} studied how regulatory focus affects children's verbal creativity using a social robot, but creativity was measured as an outcome of priming rather than as a function of the tool itself.

To summarise, we excluded papers based on the following criteria:
(1) studies reporting only formative or needs-finding work, without evaluating a CST;
(2) studies that excluded human involvement in the creative process (e.g., fully computer-generated outputs);
(3) studies focused on developing taxonomies of user groups;
(4) studies of automation with only human verification (e.g.,~\cite{yan2022flatmagic}); 
(5) studies collecting only user sentiment (e.g., likes/dislikes~\cite{watanabe2017lyrisys});
(6) studies relying solely on observational data;
(7) studies with unspecified or insufficient descriptions of outcome measures;
(8) studies where creativity was shaped primarily by cognitive prompts, rather than interactive or iterative scaffolding;
(9) studies presenting content-mining pipelines rather than interactive CSTs (e.g.,~\cite{bhavya2023cam,cox2021directed});
(10) studies focused on general interaction or perception of existing CSTs without creative outcomes;
(11) studies evaluating interventions rather than CSTs.
Unlike prior reviews that applied citation or download count thresholds~\cite{frich2018twenty,frich2019mapping}, we imposed no such restrictions in order to provide a broader overview of outcome measures.

\begin{table}[h]
\begin{tabular}{p{0.1\linewidth}p{0.65\linewidth}}
    \toprule
    \textbf{Year} & \textbf{Publication} \\
    \midrule
    2024 
& \cite{williams2024doodlebot,schlagowski2024xr,kariyawasam2024appropriate, heyman2024supermind,asha2024introducing,manesh2024sharp,weber2024wr,anderson2024homogenization, hoffmann2024thermalpen,kamihira2024multiplexed,zhang2024exploring,ning2024mimosa,peng2024designprompt,ford2024reflection,jiayu2024jamplate,wang2024podreels,yen2024give,srinivasan2024improving,chakrabarty2024creativity,oh2024lumimood,he2024interactive,davis2024fashioning,wadinambiarachchi2024effects,chen2024autospark,zhou2024understanding,yao2024lumina,choi2024creativeconnect,wang2024reelframer,chang2024editscribe,zhang2024hybrid,qin2024charactermeet,shen2024neural,wan2024metamorpheus,chen2024beyond,long2024sketchar,xiao2024typedance,fung2024create,antony2024id,heyman2024supermind,suh2024luminate,goldi2024intelligent,zhong2024ai,bi2024blow,reza2024abscribe,stemasov2024param,fan2024contextcam,john2024adaptics,wu2024stylewe,chen2024chatscratch,zhang2024protodreamer,son2024genquery,zhang2024mathemyths} \\
2023 
& \cite{chavula2023searchidea,wan2023gancollage,cai2023designaid,lawton2023drawing,pyakurel2023cogcues,kim2023metaphorian,hoque2023portrayal,chung2023artinter,hegemann2023cocolor,liu20233dall,yan2023xcreation,jeong2023automatastage,ford2023towards,yan2023nacanva,huh2023genassist,wang2023popblends,warner2023interactive,wagener2023selvreflect,angert2023spellburst,petridis2023anglekindling,guo2023creative,mirowski2023co,xu2023magical,wang2023pointshopar,wu2023styleme,jansen2023mix,park2023exergy,willett2023curvecrafter,zhang2023visar,dang2023worldsmith,singh2023hide} \\
2022 
& \cite{ali2022escape,elgarf2022creativebot,rezwana2022understanding,schlagowski2022flow,suh2022privacytoon,gero2022sparks,tamburro2022comic,zhang2022storydrawer,liu2022opal,volkmar2022effects,kim2022mixplorer,evin2022cine} \\
2021 
& \cite{elgarf2021once,hu2021exploring,ngoon2021shown,jeon2021fashionq,frich2021digital,xu2021ideaterelate,kang2021metamap,schmitt2021characterchat,lomas2021design,hwang2021ideabot,lu2021streamsketch,macneil2021framing,cho2021intumodels,sultanum2021leveraging} \\
2020 & 
\cite{buschek2020paper2wire,alves2020creativity,zimmerer2020finally,ali2020can,zarei2020investigating,koch2020semanticcollage,shi2020emog,zhu2020augmenting,bae2020spinneret,neate2020creatable,lin2020your,bourdeau2020design,shimizu2020design,koch2020imagesense,tamburro2020accessible,oppenlaender2020crowdui,wallace2020sketchy,dayama2020grids} \\
2019 & 
\cite{marinussen2019being,przybilla2019machines,ali2019can,wolf2019paint,aslan2019creativity,maiden2019evaluating,koch2019may,neate2019empowering,girotto2019crowdmuse,williford2019drawmyphoto,shugrina2019color,men2019lemo,ahmed2019structuring} \\
2018 & 
\cite{wang2018plain2fun,lundqvist2018physical,oh2018lead,wu2018thinking,fujinami2018painting} \\
2017 & 
\cite{engelman2017creativity,oberhuber2017augmented,gonccalves2017shall,gonccalves2017understanding,de2017creativity,maiden2017evaluating,willett2017secondary,girotto2017effect,ccamci2017inviso,wu2017supporting,piya2017co,kato2017f3,andolina2017crowdboard,brown2017stimulating,shugrina2017playful} \\
2016 & 
\cite{torres2016proxyprint,martin2016intelligent,chan2016improving,siangliulue2016ideahound,craveirinha2016exploring} \\
2015 & 
\cite{seehra2015handimate,de2015motion,siangliulue2015toward,de2015emotion,gonccalves2015you,kato2015textalive,siangliulue2015providing,kim2015motif} \\
    \bottomrule
\end{tabular}
\caption{List of included papers (sorted by year).}
\label{tab:year-publication}
\end{table}

\subsection{Search}
For this review, we surveyed research published in the ACM Digital Library in a 10 year time period from beginning of 2015 to end of 2024. 
When searching the ACM Digital Library, we followed prior reviews of CSTs~\cite{frich2019mapping,frich2018twenty,remy2020evaluating} by searching for ``\textit{creativity support tool}'' in any part of a publication. We also searched for papers that used ``\textit{creativity}'' as a keyword to capture studies that evaluated CSTs, but did not necessarily include the phrase in their publication.
This search resulted in a total of 1241 papers included in the initial selection process.

\subsection{Selection and Analysis}
To apply the inclusion and exclusion criteria, one researcher manually reviewed all 1241 papers, initially screening titles and abstracts, and consulting the introduction, methodology, and results sections where necessary.
During this initial screening, any exclusion-relevant criteria not specified in the original list (e.g., papers that only collected user sentiment) were noted and discussed with the wider research team.
The exclusion criteria were then revised to reflect these additions and applied throughout the remainder of the screening process.
This process resulted in 235 papers being included for outcome measure analysis.

Next, following prior work~\cite{zheng2022ux,rapp2021human} we conducted a thematic analysis of the remaining 235 papers. 
Two researchers qualitatively analysed the papers to identify and extract reported outcome measures.
An open coding process was used to extract measures, with continuous discussion between coders to ensure consistency.
Each paper was read in full to support detailed extraction of outcome measures.
\textcolor{edit}{
For example, Jamplate~\cite{jiayu2024jamplate} was coded as using NASA-TLX and SUS (among other measures) to evaluate the publication's CST.
}
In this process, we focused solely on outcome measures, and did not include any moderating or control variables (e.g., intrinsic motivation and creative thinking ability~\cite{przybilla2019machines}).

\textcolor{edit}{
During coding, an additional 62 papers were excluded, resulting in a final set of 173 papers included in the review.
These were excluded due to insufficient reporting detail (e.g., vague descriptions of ad hoc measures with no cited source) or because they met exclusion criteria upon closer examination.
}

After completing qualitative analysis for all 173 papers, the two researchers consolidated codes into themes and subthemes.
This process yielded three overarching themes: \textit{User Experience with CST}, \textit{Creative Artefact Quality}, and \textit{User-Centric Benefits}.
Codes were further organised into subthemes within each major theme.
For instance, codes such as ``\textit{SUS}'', ``\textit{UEQ}'', ``\textit{usefulness (ad hoc)}'' and ``\textit{satisfaction (ad hoc)}'' measures were grouped under the ``\textit{General Usability Measures}'' subtheme within the ``\textit{User Experience with CST}'' theme.


\textcolor{edit}{
Table~\ref{tab:year-publication} presents all 173 included papers sorted by publication year. Table~\ref{tab:all_papers} shows publications by venue, with CHI (\textit{n=55}), C\&C (\textit{n=28}), DIS (\textit{n=16}), and UIST (\textit{n=16}) as the most represented.
}

\begin{table}[h]
\begin{tabular}{p{0.8\linewidth}p{0.12\linewidth}}
    \toprule
    \textbf{Publication Venue} & \textbf{Paper Count} \\
    \midrule
    \textit{CHI}: Conf. on Human Factors in Computing Systems & 55 \\
    \textit{C\&C}: Creativity and Cognition & 28 \\
    \textit{DIS}: Designing Interactive Systems & 16 \\
    \textit{UIST}: User Interface Software and Technology & 16 \\
    \textit{CSCW}: Computer Supported Cooperative Work & 8 \\
    \textit{CHI PLAY}: Computer-Human Interaction in Play & 5 \\
    \textit{ICMI}: Int. Conf. on Multimodal Interaction & 4 \\
    \textit{IUI}: Intelligent User Interfaces & 4 \\
    \textit{CI}: Collective Intelligence & 3 \\
    \textit{TEI}: Tangible, Embedded and Embodied Interaction & 3 \\
    \textit{TOCHI}: Trans. on Computer-Human Interaction & 3 \\
    Other & 28 \\
    \bottomrule
\end{tabular}
\caption{Number of included papers (sorted by venue count). The category ``Other'' includes publication venues with two or less total publications.}
\label{tab:all_papers}
\end{table}

\textcolor{edit}{
For reader context, the reviewed literature spans a variety of application areas (though these were not part of our formal analysis). 
These include education, creative practice, industry, entertainment, and accessibility.
In education, target users range from kindergarten to university students, encompassing a broad age spectrum.
Industry-focused studies include live coders, designers, and writers, with focus on supporting ideation, design, collaboration, and content generation.
Studies on creative practice target both casual users engaging in creative activities for leisure, as well as professional artists, including musicians, painters, and creative writers.
In entertainment, target users include general audiences as well as professionals such as live-streamers and film-makers. 
Accessibility-focused work includes support for users with low vision, blindness, or aphasia.
}

\section{Findings}




We report the findings of our thematic analysis of the outcome variables contained in the \textit{n=173} papers. 
Here, we structure our findings around the three high-level themes that emerged from our thematic analysis.
Specifically, themes covered the following outcome measures:
\begin{itemize}
    \item \textbf{User Experience with CST} (Section~\ref{sec:UserExperience}): Outcomes measuring the user experience during the ideation process when interacting with the CST, such as system usability.
    \item \textbf{Creative Artefact Quality} (Section~\ref{sec:Quality}): Outcomes measuring the quality of the creative output such as the quality of creative writing, music, or artist work.
    \item \textbf{User-Centric Benefits} (Section~\ref{sec:UserCentric}): Outcomes that benefit the \textit{user themselves} outside of the CST, such as learning experiences, improvement in well-being and self-confidence.
\end{itemize}

\begin{figure*}[h]
  \centering
  \includegraphics[width=0.8\textwidth]{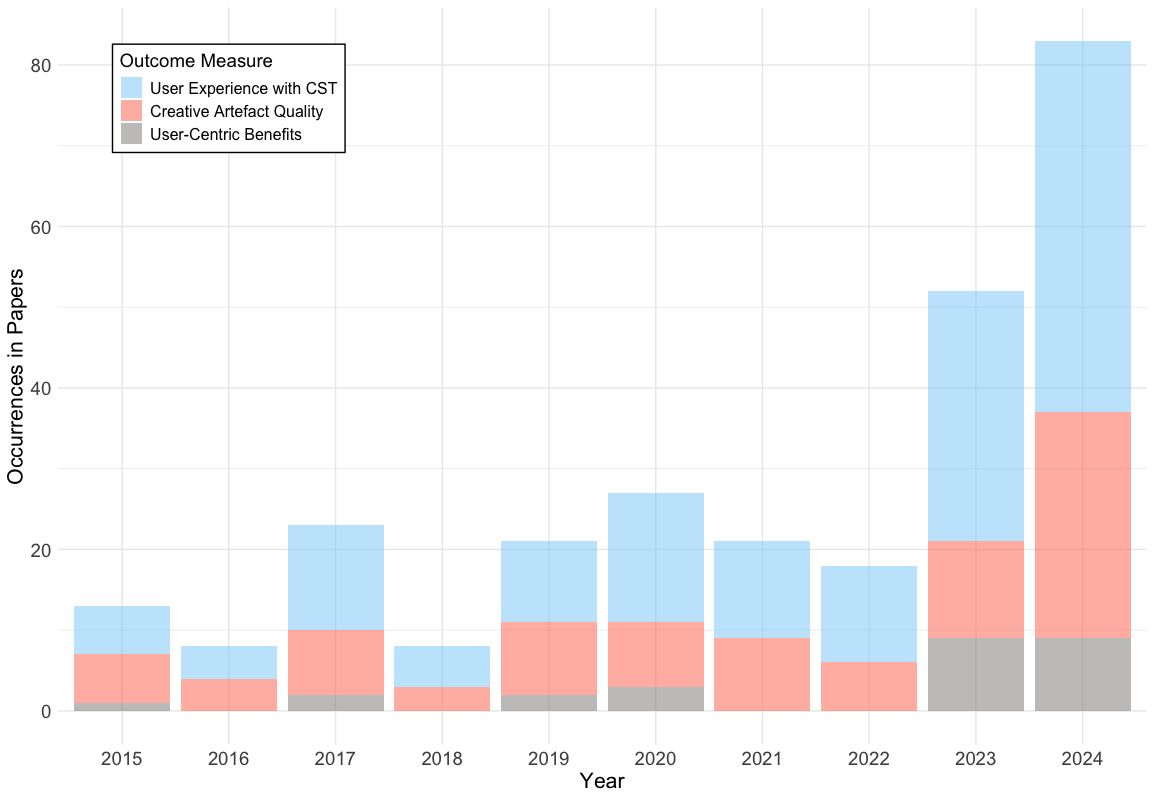}
  \caption{\textcolor{edit}{Count of measures used (\textsc{User Experience with CST}/\textsc{Creative Artefact Quality}/\textsc{User-Centric Benefits}) by year (2015–2024). As some studies use measures across multiple themes, sums will exceed total number of papers.}}
  \Description{Stacked bar chart showing yearly occurrences of three outcome measure types in CST studies from 2015 to 2024. Bars are divided into segments for User Experience (largest and consistently dominant), Creative Artefact Quality (generally second most common), and User-Centric Benefits (least frequent, with slight growth from 2019 onward). A sharp increase in all categories is visible in 2023 and 2024.}
  \label{fig:time-series}
\end{figure*}

\textcolor{edit}{Additionally, we report on the distribution of papers across sub-themes within each main theme (e.g., ``\textit{Self-Reflection}'' within the ``\textit{User-Centric Benefits}'' theme). Table~\ref{tab:Themes-Count} provides counts of papers by theme and sub-theme.
Figure~\ref{fig:time-series} shows the number of themes per year.
The rise in CST evaluations (see Fig.~\ref{fig:time-series} and Tab.~\ref{tab:year-publication}) is in part due to the increase in writing support tools and adoption of LLMs in HCI research~\cite{pang2025understanding}.
}

\begin{table}[h]
\begin{tabular}{p{0.12\linewidth}p{0.58\linewidth}p{0.12\linewidth}}
    \toprule
    \textbf{Themes} & \textbf{Sub-themes} & \textbf{Paper Count} \\
    \midrule
    \multicolumn{2}{l}{\textbf{User Experience with CST}} & \textbf{155} \\
     & General Usability Measures & 101 \\
     & Measures of Creativity Support & 58 \\
     & Interaction Time and Patterns & 52 \\
     & Collaboration with Human or AI & 36 \\
     & Workload and Cognitive Load & 26 \\
     & Engagement, Flow, and Immersion & 15 \\
    \multicolumn{2}{l}{\textbf{Creative Artefact Quality}} & \textbf{93} \\
     & Human Ratings of Quality & 74 \\
     & Quantitative Measures of Quality & 54 \\
    \multicolumn{2}{l}{\textbf{User-Centric Benefits}} & \textbf{26} \\ 
     & Developing Intrinsic Abilities & 13 \\
     & Emotional Well-being & 6 \\
     & Self-Reflection & 6 \\
     & Self-Perception & 5 \\
    \bottomrule
\end{tabular}
\caption{\textcolor{edit}{Outcome measures that emerged from the analysis. As studies use measures across multiple themes, sums will exceed total number of papers.}}
\label{tab:Themes-Count}
\end{table}


\subsection{User Experience with CST}
\label{sec:UserExperience}


Of the 173 total papers, 155 evaluated the user experience of idea generation using the CST.

\subsubsection{\textcolor{edit}{General Usability Measures:}}
Here, the majority of studies (\textit{n=101}) evaluated the general usability and user experience of the CST.
Specifically, \textit{n=84} studies incorporated ad hoc usability measures, \textit{n=20} studies adopted the System Usability Scale (SUS)~\cite{brooke1996sus}, \textit{n=4} studies adopted the Usability Questionnaire (USE)~\cite{lund2001measuring}, and \textit{n=3} studies adopted the User Experience Questionnaire (UEQ)~\cite{laugwitz2008construction}\footnote{For completeness, other validated measures used (each in two or less studies) include QUIS, SEQ, PSSUQ, and AttrakDiff.}.
Ad hoc usability measures were typically used to evaluate specific question items (for example, ad hoc ease of use and enjoyment measures for the CST~\cite{neate2020creatable}) or to evaluate the usability of individual features within the CST.
Additionally, some studies used combinations of validated measures (such as SUS) in combination with ad hoc usability measures, such as items to rate the ease of use of individual CST features (e.g.,~\cite{dayama2020grids}).

\subsubsection{\textcolor{edit}{Measures of Creativity Support:}}
Measures specifically tailored to evaluating creativity support were used by \textit{n=58} studies. Of these, \textit{n=50} studies incorporated all or partial dimensions of the Creativity Support Index (CSI)~\cite{cherry2014quantifying}\footnote{\textcolor{edit}{Note: While the CSI~\cite{cherry2014quantifying} has an Enjoyment subscale (with items such as: ``\textit{I would be happy to use this system or tool on a regular basis.}''), this is enjoyment of user experience and is therefore not coded as a user-centric outcome.}}, \textit{n=5} studies used ad hoc measures, \textit{n=2} studies used the Creativity Experience Self Rating Questionnaire (CESR)~\cite{dahl2007thinking}, \textit{n=1} study used a scale (i.e.,~\cite{mccoy2002potential}) to measure how a virtual reality environment itself fosters creativity~\cite{kamihira2024multiplexed}, and \textit{n=2} studies used the Mixed-Initiative Creativity Support Index (MICSI)~\cite{lawton2023drawing}. The MICSI adds a dimension related to human-AI co-creation to the CSI measuring the perceived control and communication with a co-creative AI partner (see Antony et al. for a creative writing CST evaluated using MICSI~\cite{antony2024id}). 

\subsubsection{\textcolor{edit}{Interaction Time and Patterns:}}
The interaction activity, time, or behaviour patterns was recorded and evaluated by \textit{n=52} studies. 
Interaction time was reported by \textit{n=37} studies with focus either framing interaction time as being more beneficial as a lower value for more efficiency (e.g.,~\cite{wu2024stylewe}) or where longer interaction time is desirable as an indicator for increased user engagement and consideration during ideation (e.g.,~\cite{wagener2023selvreflect}).
User interactions with the CST (e.g., button presses, conversational turns taken) were reported by \textit{n=31} studies, with goal of high or low interaction count following a similar motivation (i.e., either framing less interactions as desireable due to increased efficiency~\cite{warner2023interactive}, or more interactions framed as desireable due to increased user engagement~\cite{heyman2024supermind}).
Patterns of behaviour were also analysed by a number of studies~\cite{jeong2023automatastage}, and the quantity and proportion of tool function use was reported as a measure of function popularity among users (e.g.,~\cite{hegemann2023cocolor}).
Additionally, multiple studies reported the acceptance of CST suggestions (such as via semantic similarity between final artefact and CST inspirations, or as use indicated acceptance~\cite{srinivasan2024improving,hu2021exploring,ngoon2021shown}).

\subsubsection{\textcolor{edit}{Collaboration with Human or AI:}}
\textcolor{edit}{Next, \textit{n=36} studies used measures focusing on aspects of collaboration. These included measures of human-human collaboration (where a CST facilitated collaborative creativity), user perceptions of CST-generated feedback, and interactions specifically involving AI components within CSTs.}
Here, there was some crossover, with the same outcome measures being used to evaluate both human-human collaboration and human-AI co-creativity.
For example, earlier crowdsourcing or group collaboration studies often used measures of group cohesion or levels of agreement between group members (e.g.,~\cite{ahmed2019structuring,men2019lemo}). 
On from this, recent work has used similarly motivated question items to evaluate interactions with human-AI co-creativity CSTs (e.g.,~\cite{hwang2021ideabot,singh2023hide,mirowski2023co}.
However, while human-human studies may use validated scales, human-AI partnerships more often used ad hoc measures (e.g., ``\textit{I felt like I was collaborating with the AI system}''~\cite{mirowski2023co}) indicating a potential need for more research into validated measures for human-AI co-creativity.

This is particularly pressing given the rising capabilities and limitations surrounding the use of LLMs in creative work.
Here, CSTs can introduce issues surrounding hallucinations, lack of consistency and predictability, as well as perceptions surrounding perceived ownership of created artefacts and perceived agency and control when working in a human-AI partnership.
Demonstrating this breadth of potential issues and need for added measures are recent works that adopt ad hoc measures to account for emerging concerns with human-LLM interactions (e.g.,~\cite{reza2024abscribe,mirowski2023co,weber2024wr,qin2024charactermeet} where no validated measure for perceived ownership or perceived agency emerged).


\subsubsection{\textcolor{edit}{Workload and Cognitive Load:}}
Workload and cognitive load were evaluated by \textit{n=26} studies, with
\textit{n=19} studies using the NASA Task Load Index~\cite{hart1988development}, \textit{n=5} studies using ad hoc measures, and \textit{n=2} studies using other validated scales.
Here, ad hoc measures were used to generate single item measures specific to the use case of the CST. For example, Petridis et al. recruited journalists to evaluate AngleKindling, a CST to support journalistic ideation, and used the ad hoc item (``\textit{Coming up with story ideas was mentally taxing with this system.}'') to measure mental demand~\cite{petridis2023anglekindling}.

\subsubsection{\textcolor{edit}{Engagement, Flow, and Immersion Measures:}}
User engagement, flow (i.e., the feeling of losing time), and immersion was evaluated by \textit{n=15} studies.
For example, Schlagowski et al. used the Flow Short Scale to measure participants' flow while using a music production CST~\cite{schlagowski2024xr}.
Additionally, some studies used relative subjective duration (RSD) flow measure to quantify a difference between a participant's perceived interaction time and the actual interaction time (e.g.,~\cite{zimmerer2020finally,torres2016proxyprint}).
Measures of flow had crossover with scales used for video games, such as the Game Experience questionnaire, where some studies used all dimensions or flow dimension only (e.g.,~\cite{zimmerer2020finally,zarei2020investigating}). Here, only \textit{n=1} study used ad hoc measures, demonstrating the current fulfilment of validated measures.

\subsection{Creative Artefact Quality}
\label{sec:Quality}


Of the 173 total papers, 93 evaluated the quality of the artefact produced using the CST.
For example, studies evaluated the quality of innovative product designs~\cite{zhang2024hybrid}, the perceived humour of slogans~\cite{kariyawasam2024appropriate}, or the expressiveness of artwork~\cite{hoffmann2024thermalpen}.
Evaluating artefact quality generally relied on human ratings, or measures of quality derived via quantitative content analysis.
Within this theme, well established measures of creativity were generally adopted, with measures such as \textit{fluency}, \textit{flexibility}, \textit{originality} and \textit{elaboration} receiving wide use.

\subsubsection{Human Ratings of Artefact Quality:}
For human ratings, one popular approach (\textit{n=34}) was for ideators to rate the perceived quality or satisfaction of the artefact they themselves had created\footnote{As a caveat, while some studies framed self-rated satisfaction in outcome as a measure of quality of artefact~\cite{bae2020spinneret,tamburro2022comic,xu2023magical,kang2021metamap}, in other work similarly worded measures were included as part of general system usability measures. Here we took an approach to include measures as self-ratings of quality if the item explicitly referenced the artefact that was generated (e.g., ``[...] \textit{satisfied with \textbf{your painting}}''~\cite{xu2023magical}).}.
For example in Metaphorian, ideators were asked to rate the accuracy, originality and coherence of their scientific metaphors~\cite{kim2023metaphorian}, and Zhong et al. asked ideators to rate the perceived usefulness and correctness of diagrams~\cite{zhong2024ai}. Within the \textit{Creative Artefact Quality} theme, these self-rated measures were often the most ad hoc\footnote{By ``\textit{ad hoc}'' measures, we refer to measures that were generated by the researchers themselves for the purpose of their study.} with authors creating custom measures to match the particular use case of their study.

Human ratings were also provided by third parties such as experts and laypeople (\textit{n=48}). While ad hoc measures were still used, more of these ratings were structured around existing theory or techniques.
For example, Amabile's Consensual Creativity Assessment Technique~\cite{amabile1982social} was adopted by several studies to guide ratings (e.g.,~\cite{zhang2022storydrawer,williford2019drawmyphoto,chen2024chatscratch}).
Another method was to derive measures from the Torrance Thinking Creativity Test (TTCT), such as evaluations for creative writing~\cite{gonccalves2017understanding} or drawings~\cite{hoffmann2024thermalpen}. 
As a common definition of creativity is for an artefact to be both novel and feasible, these were commonly used as assessment criteria for raters who were both experts and laypeople~\cite{andolina2017crowdboard,chan2016improving,siangliulue2016ideahound,kamihira2024multiplexed,ahmed2019structuring}.



\subsubsection{Quantitative Measures of Artefact Quality:}
Additionally, artefacts were evaluated using quantitative approaches (\textit{n=54}).
Commonly CSTs were evaluated using the four indicators of divergent thinking from TTCT (i.e., fluency, flexibility, elaboration, and originality).

Here, fluency (the number of ideas) was adopted by a sizeable number of studies (\textit{n=34}). 
The precise definition of fluency within publications was similar, but with varying levels of precision. 
While some work reported fluency as the total number of ideas generated during the study, other work followed a more rigorous approach such as by excluding incomplete or irrelevant ideas~\cite{chan2016improving}, or by including only unique ideas~\cite{bi2024blow}. Further, Zhang et al. used verbal analysis methods to analysis the fluency of children's utterances during visual storytelling tasks~\cite{zhang2022storydrawer}.

Flexibility, the number of unique categories across all ideations, was used by \textit{n=13} studies (studies also defined this outcome as \textit{breadth} or \textit{variety} while applying the same high-level definition).
Flexibility was often calculated by manually coding all ideas to generate categories~\cite{anderson2024homogenization,girotto2019crowdmuse}, or by quantifying mind maps~\cite{bae2020spinneret,girotto2017effect}.
On from this, \textit{n=6} studies analysed the number of ideas per category referring to either the \textit{depth} or \textit{persistence} of ideas~\cite{bi2024blow,bae2020spinneret,przybilla2019machines,girotto2017effect,chan2016improving,kamihira2024multiplexed}.

Elaboration, the amount of detail in responses, was reported by \textit{n=15} studies.
Here, elaboration was often reported using word count~\cite{anderson2024homogenization}.
Additionally, some work analysed the content of ideas (for example the ``\textit{when, where and why}'' level of detail in stories written by children~\cite{zhang2024mathemyths}, or the level of expressivity in children's verbal storytelling utterances~\cite{elgarf2021once}).

Originality, the rarity of a response, was reported by \textit{n=15} studies. 
Here, studies calculated originality computationally (e.g.,~\cite{de2015emotion,bae2020spinneret}), by manually coding responses (e.g.,~\cite{hu2021exploring}) or by via human ratings (e.g.,~\cite{elgarf2022creativebot,hwang2021ideabot}).

Finally, \textit{n=10} studies evaluated the diversity of ideas produced.
Diversity as a concept has various definitions and methods of calculation~\cite{cox2021directed}, but can typically be conceptualised as the coverage of ideas \textit{collectively} over an idea space~\cite{siangliulue2016ideahound}.
For written artefacts, sentence embeddings were used to calculate the semantic similarity between ideas (e.g.,~\cite{anderson2024homogenization,ahmed2019structuring}). 
Here, the general motivation was to increase the diversity of ideas generated during idea exploration. While there is some overlap between diversity and the depth and breadth of ideas, diversity related studies followed a less categorical approach to quantifying ideas.



\subsection{User-Centric Benefits}
\label{sec:UserCentric}

Of the 173 total papers, 26 reported measures that benefit the \textit{user themselves} as a result of using the CST. By this, we refer to measures that are aimed towards empowering the user's intrinsic abilities, feelings, self-confidence and well-being rather than the quality of creative artefact or user experience\footnote{Note: We differentiate user-centric outcome measures from moderating or control variables. For example, intrinsic motivation and creative thinking ability was used by Przybilla et al. as a control variable and therefore not included in this theme~\cite{przybilla2019machines}.}.
User-centric measures often deployed both pre- and post-tests to measure change.
Additionally, user-centric measures were particularly prevalent within studies where children were the user group (\textit{n=10}), especially when learning outcomes were assessed.


\subsubsection{\textcolor{edit}{Developing Intrinsic Abilities:}}
\textcolor{edit}{Measures that focused on developing people's intrinsic abilities were used in \textit{n=13} studies.
Here, studies assessed outcomes such as learning of concepts and terminology, as well as improvements in creative skills, often using pre- and post-tests to evaluate change.}
Learning outcomes were measured by \textit{n=11} studies.
For example, improving AI literacy among children~\cite{williams2024doodlebot} and designers~\cite{peng2024designprompt}, helping designers learn to identify issues in visual design~\cite{yen2024give}, or learning of domain specific terminology through creative tasks~\cite{zhang2024mathemyths,jansen2023mix}.

\textcolor{edit}{
Additionally, two studies~\cite{alves2020creativity,ali2020can} used the ``Test for Creative Thinking-Drawing Production'' (TCT-DP)~\cite{jellen1986tct} to evaluate the creativity of children's drawings after using a CST. In this test, children are asked to complete an incomplete drawing within an (unknown to the children) time limit. The resultant drawings are then scored across multiple dimensions to assess creativity.
Alves-Oliveira et al. assessed children's verbal creativity using pre- and post-test divergent thinking tasks~\cite{alves2020creativity}. In these tasks, children viewed an image and verbally asked as many questions as they could, with responses being analysed to produce a single verbal creativity score.
Finally, Engelman et al.~\cite{engelman2017creativity} developed survey items based on prior creativity literature to assess traits related to creative expressiveness, exploration, immersion, and thinking skills. These items were used in pre- and post-tests to evaluate potential changes in students’ creativity.
}

\subsubsection{\textcolor{edit}{Emotional Well-being:}}

People's emotional well-being as a result of CST interaction was measured by \textit{n=6} studies\footnote{Note: While studies have examined regulating ideators' emotions to support creativity~\cite{de2015emotion,de2015motion}, these emotions are primarily tied to the user experience of the CST itself (e.g., frustration or satisfaction during use~\cite{de2015emotion}). Therefore, we do not include these studies under the theme of user-centric benefits, and instead focus on emotional outcomes that more explicitly benefit the user beyond their interaction with the CST.}.
\textcolor{edit}{
Several studies used adjective-based scales, featuring items such as ``\textit{proud}'', ``\textit{upset}'', or ``\textit{happy}''.
Participants either rated the extent to which they experienced each emotion using validated scales (such as PANAS~\cite{watson1988development} or AD-ACL~\cite{thayer1986activation}), or selected the adjectives that best described their mood (based on ad hoc measures~\cite{gonccalves2015you,gonccalves2017shall}), both before and after the interaction.
}

For example, Gonçalves et al.~\cite{gonccalves2015you} evaluated creative writing support tools with marginalised youth, measuring well-being after each daily writing session over a two-week period to track changes in well-being throughout the study.
Wagener et al.~\cite{wagener2023selvreflect} used the PANAS scale to assess participants’ emotions before and after interacting with their CST, in order to evaluate its effect on positive and negative affect.
\textcolor{edit}{Finally, Wan et al.~\cite[Table 1]{wan2024metamorpheus} designed questions using Mekler and Hornb{\ae}k’s framework for the experience of meaning~\cite{mekler2019framework}, many of which are closely tied to emotional well-being through their focus on users’ emotional responses, sense of connection, and personal resonance with the system.}

\subsubsection{\textcolor{edit}{Self-Reflection:}}

Finally, \textit{n=6} studies evaluated measures of self-reflection\footnote{Note: The term ``\textit{self-reflection}'' refers specifically to the user's \textit{personal} reflection. As such, reflection on the CST interaction itself is not included.},
with studies using both ad hoc and validated measures.
One validated measure used was the Reflection in Creative Experience (RiCE) questionnaire~\cite{ford2024reflection,ford2023towards}, a 9-item scale to measure reflection occurring in creative practice.
RiCE includes a subscale specifically focused on reflection on self, with items such as: ``\textit{I learned many new things about myself during the experience}''.
\textcolor{edit}{
Another validated measure used was the Technology-Supported Refection Inventory (TSRI)~\cite{bentvelzen2021development}, a 9-item scale to measure how well a system supports reflection. TSRI features subscales for insight, exploration, and comparison, with items such as: ``\textit{Using the system gives me ideas on how to overcome challenges}''.
}

For instance, Wagener et al. developed `\textit{SelVReflect}'~\cite{wagener2023selvreflect} to guide participants to self-reflect during an assisted drawing task, and measured self-reflection using an adjusted form of the TSRI.
Yan et al. designed `\textit{NaCanva}'~\cite{yan2023nacanva}, a CST to help children develop mood boards to aid in personal reflection (i.e., recalling memories).
`\textit{NaCanva}' also aimed to promote learning about nature, and foster a positive attitude and engagement with nature (such as improving acting and caring on nature, awareness and emotional ties to nature, and feelings of identity within nature).

\subsubsection{\textcolor{edit}{Self-Perception:}}
Measures related to the self-perception of participants were included in \textit{n=5} studies.
Specifically, studies measured participants' self-efficacy\footnote{``\textit{Self-efficacy}'' includes conceptually similar measures such as ``\textit{creative self-belief}'', and ``\textit{self-confidence}''.}
(e.g.,~\cite{chavula2023searchidea,wagener2023selvreflect,park2023exergy,engelman2017creativity}), and sense of achievement (e.g.,~\cite{chavula2023searchidea,xu2023magical}).
All studies measuring self-efficacy used both pre- and post-tests, with scales used including the General Self-Efficacy Scale (GSE)~\cite{schwarzer1995generalized} (e.g.,~\cite{wagener2023selvreflect}), and measures derived from existing scales (e.g.,~\cite{engelman2017creativity}).

For example, Park et al. developed the CST `\textit{Exergy}' to support non-experts in generating sustainable energy ideas~\cite{park2023exergy}. Perceived difficulty and technological confidence were measured pre- and post-interaction, aiming to improve users’ confidence and make sustainable solutions more accessible to non-experts.
Furthermore, Chavula et al. measured feelings of accomplishment after idea generation using CSTs~\cite{chavula2023searchidea}, and Xu et al. measured participants' sense of achievement after creating a painting~\cite{xu2023magical}.

\section{Discussion}

In this paper, we conducted a review of outcome measures used to evaluate user studies of CSTs. 
From our review, we highlight three emerging themes of outcomes, and draw attention to predominant outcomes in prior work, and highlight areas that warrant further attention from the HCI and creativity communities.
\textcolor{edit}{Specifically, we found that \textit{User Experience with CSTs} in roughly 90\% of studies, \textit{Creative Artefact Quality} was measured in roughly 54\% of studies, and \textit{User-Centric Benefits} in only 15\% of studies.}


First, we would like to contextualise this review against the rapidly growing capabilities of human-AI creativity partnerships~\cite{rezwana2023designing}, whereby LLMs (and other generative-AI methods) can simulate and replicate detailed human-like outputs~\cite{biermann2022tool}.
On from this, as creativity as a process adopts the use of AI-generated artefacts, the need for human creativity may be seen as unnecessary if one's perspective is framed around the quality of a creative artefact itself. In other words, if generative-AI can produce human-like output, then will the need for human input to creative pursuits diminish in return?
Motivated by this, we investigated the use outcome measures in CSTs historically within HCI literature, with aim to gain insight into the popularity and validity of prior used metrics.


Evaluations of user experience (UX) commonly incorporated general measures (such as understandability or ease of use), with the use of ad hoc measures being predominant.
In part, ad hoc measures were developed to evaluate UX within novel CSTs and their accompanying features.
However, such use of ad hoc measures also draws attention to the need for increased focus on creating validated measures tailored to CST evaluation. 
The use of ad hoc measures also reduces the ability to reproduce work (with some studies being insufficiently described in terms of outcome measures and therefore not suitable for inclusion in our review).

Additionally, recent CST evaluations have included the use of ad hoc measures that account for the added affordances and limitations of generative AI~\cite{hwang2021ideabot,singh2023hide,mirowski2023co}.
Alongside the rapid emergence of generative-AI for real-world creativity tasks, the use of validated measures that account for these affordances and potential limitations is needed within the creativity and HCI community.
Additionally, we would like to highlight the lack of measures adopted in studies that would be present within the conversational user interface (CUI) community (see reviews for example measures used within CUI literature~\cite{rapp2021human,zheng2022ux}).
For example, measures have been used to evaluate the quality of conversations between users and CUIs~\cite{cox2022does,grice1975logic,xiao2020if,jacobsen2025chatbots}, to evaluate user-centric benefits of conversations~\cite{kollerup2025enhancing,zhu2025benefits}, and validated scales of conversational agents have been used~\cite{song2025Stigma,wester2024ChatbotWouldNever} (such as Godspeed \cite{bartneck2009measurement,li2024stayfocused}).
As conversational interactions (such as with co-creative AI) becomes more prevalent, need for such validated measures will increase.

\textcolor{edit}{
Measures of creative artefact quality were fairly consistent and established within the literature, with measures derived from creativity literature allowing for rigour, consistency, and reproducibility.
These measures included both human ratings of creative artefacts and quantitative content analyses.
Additionally, measures were commonly derived from the Torrance Thinking Creativity Test (TTCT).
The prevalence of such TTCT related measures reflects the popularity of divergent thinking tasks within CSTs, where historically the exploration of ideas has seen increased attention compared to convergent thinking and idea refinement~\cite{frich2021digital}.
}


Finally, our review highlights a notable lack of measures that evaluate benefits to the \textit{users themselves}, 
\textcolor{edit}{such as assessments of intrinsic abilities, emotions, self-confidence, and overall well-being.}
Although recent work has highlighted a motivational shift towards user-centric outcomes (such as empowerment within artistic support tools~\cite{li2023beyond}, \textcolor{edit}{``\textit{slowing down}'' for well-being and relaxation~\cite{Falk2024SlowingDown},} or playful engagement within CSTs~\cite{liapis2023designing}) there remains a need for further research explicitly addressing these user-centric dimensions.
This need is further supported by insights from creative practitioners, who have emphasised emotional connection (e.g., feelings of happiness and belonging) as a valued aspect of CSTs~\cite{palani2022don}.
\textcolor{edit}{
The limited use of user-centric measures may come as additionally surprising given the well-established psychological and emotional benefits of creative activity~\cite{fancourt2020evidence,Curry01012005,Kaimal04052017}. The creativity literature (particularly the extensive scoping review by Fancourt and Finn~\cite{fancourt2020evidence}) offers a strong foundation to inform the design and evaluation of future CSTs in HCI.
Moreover, as behaviour change is inherently a long-term process, evaluating outcomes such as well-being or stress reduction requires longitudinal study designs in order to reliably assess the effectiveness of CSTs~\cite{kjaerup2021longitudinal,klasnja2011evaluate}.
}
In light of these motivations, we encourage greater attention to user-centric outcomes, in addition to the current emphasis on user experience and artefact quality in CST evaluation.

\subsection{Limitations}


\textcolor{edit}{
Our survey covered 10 years of literature from the ACM Digital Library. While the ACM Digital Library has been used in prior reviews as a representative sample of computing literature~\cite{zheng2022ux,frich2018twenty,frich2019mapping,remy2020evaluating}, the inclusion of additional academic repositories (such as IEEE Xplore or Science Direct) may uncover insights or measures not contained in our review, and overcome potential selection bias from the sole use of ACM publications.
Nevertheless, given that our survey systematically analysed a sample of \textit{n=173} papers spanning a 10-year period (comparable to or exceeding the scope of many prior reviews) we argue that our findings provide a robust and representative foundation for further inquiry.
}



We would like to highlight that the definition of a Creativity Support Tool (CST) is not always consistent.
On from this, we took the decision to include co-creative AI (such as LLM support~\cite{reza2024abscribe} and chatbot partners~\cite{hwang2021ideabot}).
By taking this broader approach, our review captures a wider range of tools that support creativity, ensuring a more representative analysis of the field.
Additionally, we did not include exploratory studies (e.g.,~\cite{10.1145/3173574.3174049}) in our review of outcome measures (as such studies do not include explicitly defined outcome measures by design).
\textcolor{edit}{While we acknowledge that measures of more user-centric benefits may be present in more exploratory studies, we provoke that the paucity of such measures as predefined outcomes highlights a growing need of focus for the HCI and creativity community.}

\section{Conclusion}
This paper reviewed the outcome measures used in the evaluation of CSTs as published in the ACM Digital Library, via an analysis of 173 papers as published over a ten year period (2015--2024).
Our results show that the user experience of using the CST was assessed in close to 90\% evaluated studies.
Just over half of the assessed studies explicitly evaluated the quality of the creative output.
Finally, user-centric benefits were evaluated by only 15\% of the studies included in our review.

CSTs are increasingly effective in generating both a large quantity and high quality of output, largely enabled through the use of generative AI.
This inevitably results in changes to the creative process and benefits to the users of CSTs.
We therefore urge the community to consider a more holistic approach to evaluating CSTs and carefully consider the impact of these tools on user-centric benefits.
Additionally, we highlight the need for validated measures specifically suited to the evaluation of generative AI in CSTs to overcome the current reliance on ad hoc measures.

\begin{acks}
We would like to thank our reviewers for their positive reception of the work, and their helpful and construction feedback to improve the paper.
This work is supported by the Carlsberg Foundation, grant CF21-0159.
\end{acks}

\balance
\bibliographystyle{ACM-Reference-Format}

\appendix




\end{document}